\documentclass[a4,12pt]{article}
\begin{document}
\title{Flavor Asymmetry of the Sea Quarks in the Baryon Octet}
\author{Susumu Koretune \\
Department of Physics , Shimane Medical University , \\
Izumo,Shimane 693,Japan}
\date{\today}
\maketitle
\begin{abstract}
We show that the chiral $SU(n)\otimes SU(n)$ flavor symmetry
on the null-plane severely restricts the 
sea quarks in the baryon octet. It predicts large
asymmetry for the light sea quarks $(u,d,s)$, and universality and 
abundance for the heavy sea quarks. Further it is shown
that existence of the heavy sea quarks constrained by the
same symmetry reduces the theoretical value of the 
Ellis-Jaffe sum rule substantially.
\end{abstract}
\section{Introduction}
Many years ago, based on the current anticommutation relation
on the null-plane \cite{K80}, the Gottfried sum rule 
\cite{Got} was re-derived. Since the re-derived sum rule had a 
slightly different physical meaning from the original one,
I called it as the modified Gottfried
sum rule \cite{K84}. Several years ago, this sum rule was
found to take the following form \cite{K93,K95}:
\begin{eqnarray}
\lefteqn{\int^1_0\frac{dx}{x}\{F_2^{ep}(x,Q^2)-F_2^{en}(x,Q^2)\}}& \nonumber \\
&=\frac{1}{3}\left( 1-\frac{4f_{K}^2}{\pi}\int_{m_Km_N}^{\infty}\frac{d\nu}{\nu^2}
\sqrt{\nu^2-(m_Km_N)^2}\{\sigma^{K^+n}(\nu)-\sigma^{K^+p}(\nu)\}\right) ,
\end{eqnarray}
where $\sigma^{K^+N}(\nu)$ with $N = p$ or $n$ is the total cross section of the 
$K^+N$ scatterings and $f_K$ is the kaon decay constant.
This gave us a new way to investigate the vacuum properties
of the hadron based on the chiral $SU(n)\otimes SU(n)$ 
flavor symmetry on the null-plane.  In this paper I explain
the fact and show that it severely restricts the sea quarks
in the ${\bf 8}$ baryon. Before going into details
let us first explain some backgrounds about the sum rule (1).\\
The classical moment sum rules based on the operator product
expansion(OPE) were derived only at even integers 
or at odd integers \cite{GW}, where the integer $n$ was
defined, for example, for the structure function $F_2$ as
\begin{equation}
M(n)=\int^1_0 dx x^{n-2}F_2(x,Q^2)  .
\end{equation}
Then we call the integers where the classical moment sum rules 
do not exist as the missing integers. For $F_2$ 
in the electroproduction, these missing integers are odd ones.
The reason why we lose the moments at missing integers
is clear. In the classical derivation,
we first do the short-distance expansion. Then we  
analytically continue it
to the light-cone one with use of the dispersion relation 
and expresses it by the structure function in the $s$
channel. In these processes we need some relation for the
structure function originating from the crossing 
symmetry.  We usually take this as the one defined
by the current commutation relation because we can use the
causality condition directly in this case.  Now 
this relation depends crucially on the way how
we define the quantity in the $u$ channel.
Hence by the proper consideration of this quantity,
it may be possible to get the sum rules at the missing
integers. In the perturbative QCD, the 
situation was resolved by using the analytical
continuation with respect to $n$ in the anomalous
dimension up to the two loops \cite{Ross}, and
as a by-product the Gottfried sum rule was revised as
\begin{eqnarray}
\lefteqn{\int^1_0\frac{dx}{x}\{F_2^{ep}(x,Q^2)-F_2^{en}(x,Q^2)\}}&& 
\nonumber \\
&=& \int^1_0\frac{dx}{x}\{F_2^{ep}(x,Q^2_0)-F_2^{en}(x,Q^2_0)\}
+ 0.01(\alpha_s(Q^2) - \alpha_s(Q^2_0))   .
\end{eqnarray}
Therefore if we assume 
\begin{equation}
\int^1_0\frac{dx}{x}\{F_2^{ep}(x,Q^2_0)-F_2^{en}(x,Q^2_0)\}
=\frac{1}{3} ,
\end{equation}
at some low $Q^2_0$, we get
\begin{equation}
\int^1_0\frac{dx}{x}\{F_2^{ep}(x,Q^2)-F_2^{en}(x,Q^2)\}
\sim \frac{1}{3} ,
\end{equation}
for arbitrary $Q^2$, since the $Q^2$ dependence in Eq.(3) is
negligibly small compared with 1/3. This method of the
analytical continuation in $n$ was confirmed directly
also in Ref.\cite{CFP} in the perturbative approach up to
the next to leading order. 
Now the integer $n$ was known to be the $O(4)$ spin, and
the moment $M(n)$ should be replaced by the Nachtmann
moments \cite{Nacht} which are essentially the $O(4)$ partial
waves apart from the trivial kinematical factors. Hence the 
method of the analytical continuation in $n$ could be checked from the
$O(4)$ partial wave expansion in the general context,
\cite{FW,K83}, i.e., independent of the OPE.
In this method we first defined the signatured moments
as in the case of the Froissart-Gribov projection in the
$O(3)$ partial wave expansion. Then we found that the 
$O(4)$ partial waves at the wrong signature points exactly
corresponded to the Nachtmann moments at the missing
integers.  This stemmed from the fact that the
scattering amplitude was defined by the retarded product
whose imaginary part was the commutation relation.
Now what we really needed in the inclusive reaction was
information of the product of the current. 
Thus we constructed the amplitude whose imaginary part was 
the anticommutation relation. 
By applying the $O(4)$ partial wave expansion to this
quantity, we found that the missing integers in the classical
derivation corresponded to the right signature points of
this expansion. The kinematical form of the moment at $n=1$ for 
the structure function 
$F_2$ obtained in this way in the electroproduction
was exactly the one obtained by the analytical continuation
of the method in Ref.\cite{Ross}.  However the sign from the 
crossing symmetry for the structure function was not the
usual one defined by the current commutation relation
but the one defined by the current anticommutation
relation \cite{K83}. It is this relation which is necessary to derive
the sum rule (1). A physical meaning of the difference
will be explained more in the next section.
Since a review of the derivation
of the current anticommutation relation on the null-plane 
with consideration for causality and the spectral condition 
for hadrons is described in detail in Ref.\cite{K95}, we give here the result
\begin{eqnarray}
\lefteqn{\langle p|\{J_a^+(x),J_b^+(0)\}|p \rangle_c|_{x^+=0} 
= \langle p|\{J_a^{5+}(x),J_b^{5+}(0)\}|p \rangle_c|_{x^+=0} }&&\nonumber \\
&=&\frac{1}{\pi }P(\frac{1}{x^-})\delta^2(\vec{x}^{\bot })
[d_{abc}A_c(p\cdot x , x^2=0) + f_{abc}S_c(p\cdot x , x^2=0)]p^+ ,
\end{eqnarray}
where c means to take the connected matrix element and the state
$|p \rangle $ is the stable one particle hadron state. Compared with the current
commutation relation on the null-plane given as
\begin{equation}
[J_a(x),J_b(0)]|_{x^+=0} = [J_a^5(x),J_b^5(0)]|_{x^+=0}
= if_{abc}\delta(x^-)\delta^2(\vec{x}^{\bot})J_c^+(0),
\end{equation}
the relations (6) are restricted greatly because they are not 
operator relations. In spite of this limitation we can get
many information from them.  Let us turn 
to discuss the difference between Eq.(1) and Eq.(3).
The both-hand sides of the sum rule (1) are related to the
same quantity
\begin{equation}
\frac{1}{3\pi}P\int_{-\infty}^{\infty}\frac{d\alpha}{\alpha}
A_3(\alpha ,0),
\end{equation}
where $A_a(\alpha ,0)$ on the null-plane $x^+=0$ can be considered 
as the quantity defined by 
\begin{eqnarray}
\lefteqn{\langle p|\frac{1}{2i}[:\bar{q}(x)\gamma^{\mu}\frac{1}{2}
\lambda_a q(0)-
\bar{q}(0)\gamma^{\mu}\frac{1}{2}\lambda_a q(x):]|p\rangle_c }&& \nonumber \\
&=&p^{\mu}A_a(px,x^2)+x^{\mu}\bar{A}_a(px,x^2) ,
\end{eqnarray}
as far as the moment of the structure function $F_2$ at $n=1$ is 
concerned.  We need
caution in this identification. The bilocal current on the 
left-hand side of
Eq.(9) is not the regular one in the following sense. 
Each coefficient of the
expansion of the bilocal current on the null-plane into the
local operator gets the singular piece due to the anomalous
dimension . The $Q^2$ dependence given in Eq.(3) 
corresponding to the moment at $n=1$ is one example
which originates from this fact.
However this $Q^2$ dependent piece was negligibly small.
The experimental value of the Gottfried sum \cite{NMC}
substantially violated the originally predicted 
value $1/3$ \cite{Got} and by this negligibly small value it 
became impossible to explain the defect. On the 
other hand, the modified Gottfried sum rule explained the defect \cite{K93}.  
Thus the term in the expansion
of the bilocal current corresponding to the moment at
$n=1$ should be the one predicted by our method.
In other words, we can consider
that the perturbatively predicted $Q^2$ dependence 
is shielded by the large non-perturbative effects, or
more practically, we can regard it negligible compared with
the non-perturbative contribution as far as the moment at 
$n=1$ is concerned. Thus our method
is complementary to the perturbative QCD which is believed
to be valid for the moments above $n=2$ except the large $n$
region.
Finally, it should be noted that the physical origin of the deviation 
from $1/3$ was the same as that of the axial-vector 
coupling constant from $1$ in the sense that both the deviations
are proportional to the square of the pseudo-scalar decay
constants. 

\section{The relation to the quark distribution functions}
Here we explain the role of the integral of the type 
$P\int^{\infty}_{-\infty}\frac{d\alpha }{\alpha }\cdots$ in Eq.(8).
At $x^+=0$, Eq.(8) corresponds to
\begin{equation}
\frac{1}{3\pi p^+}P\int_{-\infty}^{\infty}\frac{d\alpha}{\alpha}
\langle p|\frac{1}{2i}[:\bar{q}(x)\gamma^{+}\frac{1}{2}\lambda_3 q(0)
-\bar{q}(0)\gamma^{+}\frac{1}{2}\lambda_3 q(x):]|p\rangle_c  .
\end{equation}
We expand the quark field at $x^+=0$ as
\begin{equation}
q(x)=\sum_n a_n\phi_n^{(+)}(x) + \sum_n b_n^{\dagger}\phi_n^{(-)}(x) ,
\end{equation}
where the sum over the subscript $n$ means the spin sum and the momentum integral collectively.
Here the $(+)$ means the positive energy solution and $(-)$
the negative one.  The normal ordered product is
\begin{eqnarray}
\lefteqn{:\bar{q}(x) \gamma^+\frac{\lambda_3}{2}q(0): =
\sum_{n,m}a_n^{\dagger}a_m\bar{\phi}^{(+)}_n(x)\gamma^+\frac{\lambda_3}{2}
\phi^{(+)}_m(0)}  \nonumber \\
&\hspace{5cm}-&\sum_{n,m}b_m^{\dagger}b_n\bar{\phi}^{(-)}_n(x)
\gamma^+\frac{\lambda_3}{2}\phi^{(-)}_m(0)  .
\end{eqnarray}
The first term of the right-hand side of Eq.(12) contributes to the
quark distribution function and the second one to the antiquark 
distribution functions. Let us now define a part of the quark
distribution as
\begin{equation}
f(x)=\frac{1}{2\pi p^+}\int^{\infty}_{-\infty}d\alpha \exp [-ix\alpha ]
\langle p |\sum_{n,m}a_n^{\dagger}a_m\bar{\phi}^{(+)}_n(y^-)\gamma^+\frac{\lambda_3}{2}
\phi^{(+)}_m(0) | p \rangle_c  ,
\end{equation}
with $\alpha =p^+y^-$.
Similarly the antiquark one is
\begin{equation}
g(x)=-\frac{1}{2\pi p^+}\int^{\infty}_{-\infty}d\alpha \exp [-ix\alpha ]
\langle p |\sum_{n,m}b_n^{\dagger}b_m\bar{\phi}^{(-)}_m(0)\gamma^+\frac{\lambda_3}{2}
\phi^{(-)}_n(y^-) | p \rangle_c  .
\end{equation}
Then using the integral representation of the sign function
$\epsilon (x)=\frac{1}{i\pi}P\int^{\infty}_{-\infty}\frac{da}{a}\exp
[iax]$, we get 
\begin{eqnarray}
\lefteqn{\frac{1}{i\pi p^+}P\int^{\infty}_{-\infty}\frac{d\alpha}{\alpha}
\langle p |\sum_{n,m}a_n^{\dagger}a_m\bar{\phi}^{(+)}_n(y^-)\gamma^+\frac{\lambda_3}{2}
\phi^{(+)}_m(0) | p \rangle_c } &&\nonumber \\
&=& \int^1_0 dx \epsilon (x)f(x)
=\int^1_0 dx f(x)  ,
\end{eqnarray}
while
\begin{eqnarray}
\lefteqn{\frac{1}{i\pi p^+}P\int^{\infty}_{-\infty}\frac{d\alpha}{\alpha}
\langle p |\sum_{n,m}b_m^{\dagger}b_n\bar{\phi}^{(-)}_n(y^-)\gamma^+\frac{\lambda_3}{2}
\phi^{(-)}_m(0) | p \rangle_c }&& \nonumber \\
&=& -\int^1_0 dx \epsilon (-x)g(x)
= \int^1_0 dx g(x)  ,
\end{eqnarray}
Thus we see that the antiquark term gets extra minus sign due to the
integral of the type $P\int^{\infty}_{-\infty}\frac{d\alpha }{\alpha}\cdots$.
This explains a physical difference between the modified Gottfried sum rule
and the Adler sum rule.  In the parton model we know 
\begin{eqnarray}
\lefteqn{\frac{3}{2}\int^1_0\frac{dx}{x}\{F_2^{ep}(x,Q^2) - F_2^{en}(x,Q^2)\}} 
&\nonumber \\
&=[\int^1_0dx\{\frac{1}{2}u_v - \frac{1}{2}d_v\} + \int^1_0dx\{\frac{1}{2}
\lambda_u - \frac{1}{2}\lambda_d\}] - [\int^1_0dx\{-\frac{1}{2}
\lambda_{\bar{u}} + \frac{1}{2}\lambda_{\bar{d}}\}]  ,
\end{eqnarray}
and 
\begin{eqnarray}
\lefteqn{\frac{1}{4}\int^1_0\frac{dx}{x}\{F_2^{\bar{\nu}p}(x,Q^2) - F_2^{\nu p}(x,Q^2)\}}
& \nonumber \\
&=[\int^1_0dx\{\frac{1}{2}u_v - \frac{1}{2}d_v\} + \int^1_0dx\{\frac{1}{2}
\lambda_u - \frac{1}{2}\lambda_d\}] + [\int^1_0dx\{-\frac{1}{2}
\lambda_{\bar{u}} + \frac{1}{2}\lambda_{\bar{d}}\}]   ,
\end{eqnarray}
where the subscript $v$ means the valence quark and $\lambda_i$ means the 
$i$ type sea quark. The difference between the two sum rules is the
sign in front of the antiquark distribution. The current
anticommutation relation on the null-plane which leads to
the modified Gottfried sum rule is proportional to
$P\frac{1}{x^-}$. This factor is the origin
of the integral of the type 
$P\int^{\infty}_{-\infty}\frac{d\alpha }{\alpha }\cdots$.
While the current commutation relation
on the null-plane which leads to the Adler sum rule is 
proportional to $\delta(x^-)$. 
Thus, under the assumption $\int^1_0dx\lambda_i(x,Q^2)=
\int^1_0dx\lambda_{\bar{i}}(x,Q^2)$, the Adler sum rule measures
the mean $I_3$ of the \{[quark] + [antiquark]\} and hence the
one of the valence quarks being equal to the $I_3$ of the proton,
while the modified Gottfried sum rule measures the 
mean $I_3$ of the \{[quark] - [antiquark]\}in the proton.
In conclusion we can say that the modified Gottfried sum rule 
naturally explains the physical difference between the
Adler sum rule and the Gottfried sum and that it
directly probes the vacuum of the proton. This fact is the
fundamental importance of the sum rules obtained by
the current anticommutation relations (6).

\section{The symmetry constraint on the light sea quark
distributions in the baryon octet}
Many years ago, Weinberg showed that at high energy there is a
symmetry closely connected with the dynamic symmetry at low
energy \cite{Wein}. The pion coupling matrix discussed there is
very similar to the matrix element $<\beta |J_a^{5+}(0)|\alpha >$
on the null-plane at $q^+=0$ where $q=p_{\beta }-p_{\alpha }$
and $\alpha , \beta$ are one-particle hadron states. This is because
this matrix element picks up the non-pole term since
$<\beta |J_a^{5\mu }(0)|\alpha > = \{g^{\mu \nu } -
q^{\mu}q^{\nu}/(q^2-m_{\pi}^2)\}<\beta |J_{a\nu}^{5}(0)|\alpha >_N$
where $N$ means the non-pole matrix element. Because of this 
property the light-like charge algebra plays very similar role
as the algebra of the pion coupling matrix. It relates the low 
energy to the high energy and the relation between the two 
energy regions is controlled by the symmetry. 
It is this symmetry which we discuss in this paper.\\
Now $A_a(\alpha ,0)$ is governed by this symmetry,i.e.,the
chiral $SU(3)\otimes SU(3)$ flavor symmetry on the null-plane. 
If we take the state on the left-hand side of Eq.(9)
as the ${\bf 8}$ baryon, it becomes
\begin{eqnarray}
\lefteqn{\langle \alpha , p|\frac{1}{2i}
[:\bar{q}(x)\gamma^{\mu}\frac{1}{2}\lambda_a q(0)
-\bar{q}(0)\gamma^{\mu}\frac{1}{2}\lambda_a q(x):]|\beta ,p\rangle_c} && \nonumber \\
&=&p^{\mu}(A_a(px,x^2))_{\alpha \beta }+x^{\mu}(\bar{A}_a(px,x^2))_{\alpha \beta },
\end{eqnarray}
where $\alpha , \beta $ are the symmetry index 
specifying each member of the ${\bf 8}$ baryon.  
Since the matrix element can be classified by the flavor singlet in the
product ${\bf 8\otimes 8\otimes 8}$, $(A_a(\alpha ,0))_{\alpha  \beta }$ 
is decomposed as
\begin{equation}
(A_a(\alpha , 0))_{\alpha  \beta } = if_{\alpha a \beta}F(\alpha,0)
+ d_{\alpha  a \beta }D(\alpha,0)
\end{equation}
for $a\neq 0$.  Using the value of the modified Gottfried
sum rule estimated in Ref.\cite{K93}, we obtain
\begin{eqnarray}
\frac{1}{3\pi}P\int_{-\infty}^{\infty}\frac{d\alpha}{\alpha}
A_3^p(\alpha ,0)& = & 
\frac{1}{3\pi}P\int_{-\infty}^{\infty}\frac{d\alpha}{\alpha}
\{ F(\alpha,0) + D(\alpha , 0)\} \nonumber \\
&=& 0.26 . 
\end{eqnarray}
The mean hypercharge sum rule in Ref.\cite{Ka} gives us
\begin{eqnarray}
\frac{\sqrt{3}}{3\pi}P\int_{-\infty}^{\infty}\frac{d\alpha}{\alpha}
A_8^p(\alpha ,0)& = & 
\frac{\sqrt{3}}{3\pi}P\int_{-\infty}^{\infty}\frac{d\alpha}{\alpha}
\sqrt{3}\{ F(\alpha,0) - \frac{1}{3}D(\alpha , 0)\} \nonumber \\
&=&2.12 .
\end{eqnarray}
Here we use the notation $A_a^B(\alpha ,0)$ with 
$B = ( p , n , \Sigma^{\pm} ,\Sigma^0 , \Lambda^0 ,
\Xi^- , \Xi^0 )$ to specify each member of the ${\bf 8}$
baryon. From Eqs.(21) and (22), we obtain 
\begin{equation}
\widetilde{F} \equiv\frac{1}{2\pi} P\int^{\infty}_{-\infty}
\frac{d\alpha}{\alpha}F(\alpha ,0) = 0.89 ,
\end{equation}
\begin{equation}
\widetilde{D} \equiv\frac{1}{2\pi} P\int^{\infty}_{-\infty}
\frac{d\alpha}{\alpha}D(\alpha ,0) = -0.50 .
\end{equation}
These are the constraints by the chiral flavor symmetry
and the discussion can be extended to $SU(4)\otimes SU(4)$ 
or still higher symmetry with some restrictions explained later.\\
Let us continue to discuss the $SU(3)$ case. 
With use of $\widetilde{F}$ and $\widetilde{D}$ 
given in Eqs.(23) and (24), it is possible to get the constraints
on the sea quarks in each baryon by repeating the same kind 
of the discussions as those for the proton \cite{K95}. We first
regularize the sum rule by using analytical continuation
from the non-forward direction. This regularization is 
based on the method in Ref.\cite{Alwis} and its application
to our case was explained in detail in Ref.\cite{K84}.
The important point in this method lies in the fact
that the sum rule is convergent under the physically reasonable
assumption,i.e., the trajectory of the pomeron satisfies 
$\alpha_P(t) < 1$ for some small $t$ where $t$ is the 
momentum transfer.  Thus once we can identify the parts 
which become divergent as $t$ goes
to zero, we can safely subtract them from both-hand sides of the
sum rule.  The soft pomeron by Donnachi and
Landshoff \cite{DL} is one example which makes it possible
to carry out the program easily. Now the assumption $\alpha_P(t) < 1$ 
for some small $t$ can not be satisfied by the hard pomeron based on
the fixed-coupling constant \cite{hard}.
However, there are great efforts to improve the defect
of this pomeron \cite{For}. The next-to-leading
corrections seems to suggest a substantial reduction of the
value of the intercept  \cite{Cia}.
The multiple scatterings of the pomeron gives us important
unitary corrections at low $x$ \cite{Mue}. Thus even in such
a perturbative approach there is a hope to satisfy the
assumption. We use the soft pomeron to explain the 
regularization, but in view of the situation, 
we clarify the quantities which do not depend on the
assumed high energy behavior in the
following.  
Now the discussion of the regularization in the non-forward
direction is cumbersome kinematically, and the technical
aspect of the method can be explained by the effective
method in the forward direction, hence 
we recapitulate it here. However, It should be
noted that, corresponding to the finite sum rules 
in this effective method, there 
always exist truly convergent sum rules in the above sense.
The sum rule for $F_2^{ep}$ in $SU(3)$ is
\begin{equation}
\int^1_0\frac{dx}{x}F_2^{ep} =\frac{1}{18\pi}
P\int^{\infty}_{-\infty}\frac{d\alpha}{\alpha}\{
2\sqrt{6}A_0(\alpha ,0) + 3A_3^p(\alpha ,0) +
\sqrt{3}A_8^p(\alpha ,0)\} .
\end{equation}
We take the leading high energy behavior of $F_2^{ep}$
is given by the pomeron as $(\frac{1}{Q^2})^{\alpha_P(0)-1}
\beta_{ep}(Q^2,1-\alpha_P(0))(2\nu )^{\alpha_P(0)-1}$,
and assume it to be the flavor singlet, where
$\alpha_P(0)$ is the intercept of the pomeron.  
Note that what we assume here is only the high energy behavior
$(2\nu )^{\alpha_P(0)-1}$ and no assumption is made about the
$Q^2$ dependence, since all the unknown $Q^2$ dependence
is absorbed in $\beta_{ep}$. This also applies to the
scale factor in $2\nu$.
Then the regularization of the sum rule goes as follows.
We rewrite the left-hand side of Eq.(25) as
\begin{eqnarray}
\lefteqn{\int_0^1\frac{dx}{x}F_2^{ep} = \int_0^1\frac{dx}{x}
\{F_2^{ep} - \beta_{ep}(Q^2,1-\alpha_P(0))x^{1-\alpha_P(0)}
\}}&& \nonumber \\
&\hspace{5cm}+& \int_0^1\ dx \beta_{ep}(Q^2,1-\alpha_P(0))x^{-\alpha_P(0)} ,
\end{eqnarray}
and, since the pomeron term is assumed to be flavor singlet,
we rewrite the right-hand side of it as
\begin{eqnarray}
\lefteqn{\frac{\sqrt{6}}{9\pi}P\int_{-\infty}^{\infty}\frac{d\alpha}{\alpha}
A_0(\alpha ,0)}&& \nonumber \\
& =&\frac{\sqrt{6}}{9\pi}P\int_{-\infty}^{\infty}
\frac{d\alpha}{\alpha}\{A_0(\alpha ,0) - f(\alpha)\} 
+\frac{\sqrt{6}}{9\pi}P\int_{-\infty}^{\infty}\frac{d\alpha}{\alpha}
f(\alpha ) .
\end{eqnarray}
By setting $\alpha_P(0)=1+b-\epsilon$, we expand $\beta_{ep}$
as $\beta_{ep}^0(Q^2) - (\epsilon -b)\beta_{ep}^1(Q^2) + 
O((\epsilon - b)^2)$. \footnote{ Note that the definition of $\beta_{ep}^1$ in
Ref.\cite{K93} is different from the $\beta_{ep}^1$ in this paper
in its sign.} The pole term as $\epsilon \to b$ should
be canceled out from both-hand sides of Eq.(25) since the sum
rules are convergent for the arbitrary finite positive
$(\epsilon -b)$ which
corresponds to the small negative $t$ in the non-forward case, hence there
must exists $f(\alpha )$ such that the quantity 
$\frac{2\sqrt{6}}{9}\widetilde{f}(\alpha )
= (\frac{2\sqrt{6}}{9}\frac{1}{2\pi}P\int_{-\infty}^{\infty}
\frac{d\alpha}{\alpha}f(\alpha ) - \frac{\beta^0_{ep}}{\epsilon -b})$
becomes finite in the limit $\epsilon \to b$,
where $\beta_{ep}^0$ is $Q^2$ independent since Eq.(25) holds at any
$Q^2$. After taking out the singular piece we take the limit 
$\epsilon \to 0$ and obtain 
\begin{eqnarray}
\lefteqn{\int^1_0\frac{dx}{x}\{F_2^{ep}-\beta_{ep}^0x^{-b}\}}&&\nonumber \\ 
&=& \frac{1}{18\pi}
P\int^{\infty}_{-\infty}\frac{d\alpha}{\alpha}\{
2\sqrt{6}S_0^3(\alpha ,0,Q^2) + 3A_3^p(\alpha ,0) +
\sqrt{3}A_8^p(\alpha ,0)\} ,
\end{eqnarray}
where $S_0^3(\alpha ,0,Q^2)$ and $\widetilde{S}_0^3$ are defined as
\begin{eqnarray}
\frac{2\sqrt{6}}{9}\widetilde{S}_0^3 &=& 
\frac{2\sqrt{6}}{9}[\widetilde{f}(\alpha ) + 
\frac{1}{2\pi}P\int_{-\infty}^{\infty}\frac{d\alpha}{\alpha}
\{{A}_0(\alpha ,0)-f(\alpha)\}] + \beta_{ep}^1(Q^2)  \nonumber \\
&=& \frac{2\sqrt{6}}{9}\frac{1}{2\pi}P\int_{-\infty}^{\infty}
\frac{d\alpha}{\alpha}S_0^3(\alpha ,0,Q^2) .
\end{eqnarray}
Here the superscript 3 in $S_0^3(\alpha ,0,Q^2)$ and
$\widetilde{S}_0^3$ means the singlet in $SU(3)$.
By comparing Eqs.(25) and (28) we see that 
the regularization of Eq. (25) simply results in 
the $Q^2$ dependence in the singlet component 
and that all the relation from the symmetry is inherited.
By keeping this fact in mind,
We define the sea quark distribution 
of the $i$ type one in the ${\bf 8}$ baryon as $\lambda^B_i$, and
regularizes its mean number as
\begin{equation}
\langle \widetilde{\lambda}^B_i\rangle = \int_0^1dx \{
\lambda_i^B - ax^{-\alpha_P(0)}\} ,
\end{equation}
where $\alpha_P(0)$ here is $\alpha_P(0)=1+b$ and is taken as
$1.0808$\cite{DL}, and $a$ is obtained as
\begin{equation}
\lim_{x\to 0}x^{\alpha_P(0)}\lambda_i^B =a .
\end{equation}
Note that the superscript $B$ drops out in $a$.
This means that the result holds for all members in
the ${\bf 8}$ baryon. In the
following we use the policy to drop the superscript B 
in the case where the relation holds for all members 
in the ${\bf 8}$ baryon. Further the suffix $i$ is dropped,
since the value $a$ is proportional to $\beta_{ep}^0$ and was 
determined through the sum rule as $a=1.2$ for $i=u,d,s$
in Ref.\cite{K95}.  This universality is the result
in our formalism obtained from the assumption that the pomeron 
is flavor singlet . This can be seen by the fact that
all the coefficients of the singlet component in
the sea quark distributions are the same.
Now, in general we can take the high energy behavior more complicated
than $\nu^{\alpha_P(t)-1}$. For example in case of
$(\ln \nu )^n\nu^{\alpha_P(t)-1}$, we will obtain the expansion
of the form $[a_1/(\epsilon -b)^{n+1} + a_2/(\epsilon -b)^{n}
+ \cdots + a_n/(\epsilon -b)]$ in stead of the simple pole
$\beta_{ep}^0/(\epsilon -b)$. All these terms belong to
the singlet and hence controlled by the symmetry. Thus ambiguities
due to high energy behavior can be absorbed into the 
finite part as $S_0^3(\alpha ,0,Q^2)$ also
in this case. Eqs.(30) and (31) should be modified
appropriately according to the assumed high energy behavior. 
Though we lose the explicit value
$a=1.2$ in this case, we still have the relation which
relates the coefficient of the leading singularity
in $F_2$ and that in the
pion nucleon reactions through the sum rule. 
Now returning back to the soft pomeron we explain how the constraints
on the sea quark distributions are obtained. 
As an example let us take the proton matrix element in Eq.(19)
with $K=\frac{\lambda_a}{2} = 
diag(1\; 0\; 0)=\frac{\sqrt{6}}{6}\lambda_0
+ \frac{1}{2}\lambda_3 + \frac{\sqrt{3}}{6}\lambda_8$,
where $diag(a\; b\; c)$ means the $3\times 3$ matrix whose
diagonal element is $a,b,c$ and all off-diagonal elements
are zero .  For the proton $\alpha = \frac{1}{2}(4 + i5) ,
\beta = \frac{1}{2}(4 - i5) $.  Then by taking 
$<\widetilde{\lambda}^B_i> = <\widetilde{\lambda}^B_{\bar{i}}>$,
we obtain $<u_v^p> + 2<\widetilde{\lambda}^p_u> =
\frac{\sqrt{6}}{3}\widetilde{S}^3_0 + 2\widetilde{F}
+ \frac{2}{3}\widetilde{D}$. Since it is straightforward to repeat
the same kind of the calculation for other sea quarks, 
we give the results in Table \ref{T1},
where we use $\widetilde{S}^3$ defined as $\widetilde{S}^3 =
\frac{\sqrt{6}}{3}\widetilde{S}^3_0$. \footnote{If we 
set $\widetilde{S}^3 = \frac{9}{10}\beta
+ 1.53$, the results for the proton in Ref.\cite{K95} 
can be reproduced.}
By using the fact that each valence part is merely the number 
of the valence quark, we get many sum rules from the relations in
Table \ref{T1}. Among them the sum rules for the mean quantum numbers 
of the light sea quarks are fundamental since they do not depend on 
$\widetilde{S}^3$. Here the light sea quarks
mean the $u,d,s$ type sea quarks. We summarize them in Table \ref{T2}. 
Note that $\widetilde{\lambda}^B_i$ is replaced by
$\lambda^B_i$ because the divergent part is canceled out
in each expression through the condition (31). 
The same kind of the fact holds also in the general case
as far as the divergent part is assumed to be flavor singlet.
Practically, all the results in Table \ref{T2} can be obtained
by considering the quantity corresponding to $\langle I_3 \rangle$,
$\langle Y \rangle$, and $\langle Q \rangle$ for the light
sea quarks directly as in the example in the section 2.
In this case the singlet component do not appear explicitly,
hence if we assume that the pomeron is flavor singlet and that
it contributes universally to every sea quark, we do not
encounter the divergence. Though such a practical approach
obscures the physical view that at high energy there is
a symmetry closely connected with the dynamic symmetry
at low energy, it convinces us that the results in
Table \ref{T2} are insensitive to the regularization.
The perturbative QCD corrections to
these relations begin from the 2 loops and 
they enter the same way as the one in the  modified
Gottfried sum rule. Therefore they are negligibly small
compared with the non-perturbative values listed
in Table \ref{T2}. 

\section{The symmetry constraint on the heavy sea quark
distributions in the octet baryon }
Here we extend the symmetry from $SU(3)\otimes SU(3)$
to $SU(n)\otimes SU(n)$ with $n \ge 4$. In general,
such symmetry is considered to be badly broken in the
Hamiltonian.  However, in our case we do not use it
explicitly. Our starting point is the local current algebra
on the null-plane for good-good component. In this case
we do not need the equation of motion, hence in this sense 
all the results in this paper do not depend on the Hamiltonian 
which badly breaks the symmetry. However we need the assumption
concerning the classification of the matrix elements by the symmetry.
It is not clear how far such classification holds.
We know that the method works well for the
badly broken $SU(3)$.  Then it may be useful to have physical 
predictions under this extension.  As it is explained later in this
section and in the next section the physical results obtained 
seem to have some experimental relevance.  
Let us now discuss the heavy sea quarks in the ${\bf 8 }$
baryon. For concreteness we take the chiral $SU(4)\otimes SU(4)$ 
flavor symmetry. In this case the ${\bf 8 }$ baryon
belongs to ${\bf 20_M}$, and the currents to 
${\bf 15}$. The matrix element in this case can be
classified by the singlet component in the product
${\bf \bar{20}_M \otimes 20_M \otimes 15}$.
Since the adjoint representation ${\bf 15}$ 
appears twice in the product as ${\bf \bar{20}_M \otimes 20_M 
= 175 \oplus 84 \oplus 45 \oplus \bar{45} \oplus 20 \oplus
15 \oplus 15 \oplus 1}$,  and since only these two ${\bf 15}$ can 
make the singlet with the remaining ${\bf 15}$,
we have two different terms in the matrix element.
Further these two ${\bf 15}$ can be represented
by the $4 \times 4$ matrix whose $3 \times 3$
sub-matrix agrees with the $3 \times 3$ matrix in 
$SU(3)$. Now two different terms have already
appeared in Eq.(20) as $F(\alpha , 0)$ and $D(\alpha , 0)$ 
in $SU(3)$. Hence, even in $SU(4)$, 
Eq.(20) holds at least for $\alpha ,\beta = 1 \sim 8$
and $a = 1 \sim 15$. However, in this generalization,
the singlet in $SU(3)$ is not the singlet
in $SU(4)$. To see this fact more concretely, we
take the matrix $K = diag(1\; 0\; 0\; 0)$, 
and decomposes it as $K = \frac{\sqrt{2}}{4}\lambda_0^4
+ \frac{1}{2}\lambda_3^4 +\frac{\sqrt{3}}{6}\lambda_8^4
+ \frac{\sqrt{6}}{12}\lambda_{15}^4$. Here the $\lambda_k^4$ 
is the Gell-Mann matrix generalized to $SU(4)$. $SU(3)$
singlet part in this decomposition is
$\frac{\sqrt{2}}{4}\lambda_0^4 + \frac{\sqrt{6}}{12}\lambda_{15}^4
= \frac{1}{3}diag(1\; 1\; 1\; 0)$. Since the 
$3\times 3$ sub-matrix $diag(1\; 1\; 1)$ is expressed as
$\frac{\sqrt{6}}{2}\lambda_0$ in $SU(3)$, the coefficient
of the singlet part in $SU(3)$ is different from the one in $SU(4)$.
On the other hand, $3\times 3$ sub-matrix in the part
$\frac{1}{2}\lambda_3^4 + \frac{\sqrt{3}}{6}\lambda_8^4$
has the same expression in these two cases. Thus we find
one relation between the singlet contribution in $SU(3)$
and the one in $SU(4)$. If we denote the $SU(4)$ singlet 
contribution as $\widetilde{S}^4_0$ corresponding to  
$\widetilde{S}^3_0$ in $SU(3)$, we obtain 
$\frac{\sqrt{6}}{3}\widetilde{S}^3_0
= \frac{\sqrt{2}}{2}\widetilde{S}^4_0 + \frac{1}{3}\widetilde{D}$.
Expressed in the parton model,
this generalization from $SU(3)$ to $SU(4)$ corresponds
to the addition of the charm sea quark with the 
condition (31) without changing
anything in the light sea quarks.  After all we find
that all the relations in Table \ref{T1} and Table \ref{T2} hold
in $SU(4)$ without changing anything.  The discussion
here also shows that all the charm sea quark distributions in the
${\bf 8}$ baryon corresponding
to the matrix element of $diag(0\; 0\; 0\; 1)$ are the
same . Explicitly in case of the soft pomeron 
we obtain $2<\widetilde{\lambda}_c> =
\widetilde{S}^3 - \frac{4}{3}\widetilde{D}$ for all  
members in the ${\bf 8}$ baryon. Since we already
have the result that the charm sea quark in the proton
is abundant \cite{K95,K92}, we reach the conclusion
that the charm sea quark is universal and abundant in 
the ${\bf 8}$ baryon. It should be noted that the gluon 
fusion like term is in general included in our definition of the 
charm quark distribution function, but the gluon
in our case is not necessarily the perturbative one.
Experimentally HERA found \cite{H1} 
abundance of the charm sea quark in the proton 
which is qualitatively similar to the one 
investigated in the toy model in Ref.\cite{K95} in the following
two respects. The one is that abundance of the charm sea
quark is correlated with the rapid rise of the structure
function $F_2$ in the small $x$ region. 
The other is that this rise persists even at small $Q^2$, which
can naturally be understood by this abundance.
Further some unpleasant features of this toy model which come
from the constraint obtained by the soft pomeron analysis may be
improved if we take into account the perturbative analysis. However,
even in such a case the property of abundance of the charm
sea quark in the above sense remains.
Now the mean quantum number $<Y>$ and $<Q>$ extended
to $SU(4)$ are not so useful since they do not
correspond to the traceless matrix. Hence the 
dependence on $\widetilde{S}_0^4$ remains. Rather
even in $SU(4)$, the mean quantum numbers of the
light sea quarks in Table \ref{T2} are useful.
The heavy quark effect should be examined by the mean
quantum number such as $<\lambda_c - \lambda_s>$,
which corresponds to the sum of the mean charm and
the mean strangeness. \\
The same kind of the discussions can be repeated in the 
$SU(5)$ or $SU(6)$, and we get 
$ 2<\widetilde{\lambda}_b> =2<\widetilde{\lambda}_t> 
= \widetilde{S}^3 - \frac{4}{3}\widetilde{D} $ 
with the constraint (31)
for the bottom and the top sea quarks in case of the soft pomeron.

\section{Flavor asymmetry of the spin-dependent sea quark
distribution}
It is interesting to note that similar discussion
to extend the symmetry from $SU(3)$ to
$SU(4)$ can be applied to the matrix element
$<p,s,\alpha | J^{5\mu}_a(0)|p,s,\beta >$ which is
directly related to the Ellis-Jaffe sum rule\cite{Ellis}.
Let us first discuss the $SU(3)$ case.  We define
\begin{equation}
\langle p,s,\alpha |J_a^{5\mu}(0)|p,s,\beta \rangle = s^{\mu}A_a^{\alpha
\beta}  ,
\end{equation}
where $s^{\mu}$ is the spin vector, and 
\begin{equation}
A_a^{\alpha \beta }=if_{\alpha a \beta }F + d_{\alpha a \beta }D ,
\end{equation}
for $a \neq 0$ . The Ellis-Jaffe sum for the ${\bf 8}$ baryon is
\begin{equation}
I_f^B=\int_0^1 dxg_1^{B}(x,Q^2) ,
\end{equation}
where the subscript $f$ specifies the flavor group.
$I_f^B$ is proportional to $d_{\alpha a \beta}$ and in case
of the proton it is well known to take the form 
\begin{equation}
I_3^p = \frac{1}{36}[4\triangle Q_0^p + 3\triangle Q_3^p
+\triangle Q_8^p ] ,
\end{equation}
where $\triangle Q_0^p =\triangle u^p + \triangle d^p + \triangle s^p $,
$\triangle Q_3^p=\triangle u^p - \triangle d^p$, and $\triangle Q_8^p = 
\triangle u^p + \triangle d^p - 2\triangle s^p $, and $\triangle q^p$
is the fraction of the spin of the proton carried by the spin
of quarks of flavor $q$. Here $\triangle q^p$ includes the 
contribution from the antiquark as usual.
Since $\triangle Q_a^p$ is proportional to $A_a^{\alpha \beta}$,
We can apply Eq.(33) to this quantity. Thus we obtain 
\begin{equation}
\triangle u^p=\frac{1}{3}S + \frac{1}{3}D +F ,
\triangle d^p=\frac{1}{3}S - \frac{2}{3}D ,
\triangle s^p=\frac{1}{3}S + \frac{1}{3}D - F, 
\end{equation}
where $\triangle Q_0^p=S$.  It is straightforward to get
the spin fraction of the quarks in other baryons.
We summarize the result in Table \ref{T3}. 
Now in $SU(4)$, $\triangle Q_{15}^B$ can be defined as
\begin{equation}
\triangle Q_{15}^B=\sqrt{6}A_{15}^{\alpha \beta}=\triangle u^B 
+ \triangle d^B + \triangle s^B - 3\triangle c^B .
\end{equation}
Applying Eq.(33) to this quantity we obtain 
$\triangle Q_{15}^B = 2D$ for all members in the ${\bf 8 }$ baryon.
Since $\triangle Q_{a}^B$ for $1\le a \le 8$ is the same as 
in $SU(3)$, we obtain 
\begin{equation}
\triangle c=\frac{1}{3}S - \frac{2}{3}D ,
\end{equation}
for all members in the ${\bf 8 }$ baryon. Note that we use the same
$S$ as in $SU(3)$. Thus we get
\begin{eqnarray}
\lefteqn{I_4^p=\frac{1}{2}[\frac{4}{9}\triangle u^p 
+ \frac{1}{9}\triangle d^p + \frac{1}{9}\triangle s^p
+ \frac{4}{9}\triangle c^p ]} \nonumber \\
&=& \frac{5}{27}S + \frac{1}{6}F -\frac{5}{54}D .
\end{eqnarray}
The generalization to $SU(5)$ or $SU(6)$ is 
straightforward , and we obtain
\begin{equation}
\triangle b = \triangle t = \frac{1}{3}S - \frac{2}{3}D .
\end{equation}
Using experimental value of $F=0.46\pm 0.01$ and $D=0.79\pm 0.01$
\cite{Hsu}, we see that for a reasonable value of $S$,
the theoretical value of the Ellis-Jaffe sum rule is reduced
substantially by the charm quark.  It is usually considered that
the light sea quark gets contribution from the
gluon anomaly because of the small-ness of the quark mass
compared with the infra-red cutoff \cite{Mank}.
The magnitude of this gluon contribution is determined 
by input information. Then, to make the Ellis-Jaffe
sum rule consistent with the experiment by
this gluon polarization, it must be taken very large.
In our case, such large gluon polarization is not
necessary.  The heavy quark such as the charm one
is suffice to make it consistent with the experiment.

\section{Conclusions}
In conclusion we show that the chiral
$SU(n)\otimes SU(n)$ flavor symmetry on the null-plane
combined with the fixed-mass sum rule developed
in Refs.\cite{K80,K84,K93,K95,Ka,K92} severely
restricts the sea quark in the ${\bf 8}$
baryon. It predicts a large asymmetry for the
light sea quarks, and universality and abundance
for the heavy sea quarks. Further we show
that the same symmetry restricts the fraction of the spin 
of the ${\bf 8}$ baryon carried by the quark.
Especially We show that 
this effect is outstanding for the intrinsic 
charm sea quark in the nucleon and that it plays the 
role to reduce the theoretical value of 
the Ellis-Jaffe sum rule substantially.

\newpage
\renewcommand{\arraystretch}{1.2}
\begin{table}
\caption{The regularized mean sea quark number. 
Here the mean valence quark number is nothing but
the valence quark number, hence it takes the value 0 or 1 or 2 
according to the valence contents of the baryon.
Further we set $\widetilde{S}^3=\frac{\sqrt{6}}{3}\widetilde{S}^3_0$.
\label{T1}}
\begin{tabular}{llll}
$B$ & $<u_v^B> +  2<\widetilde{\lambda}^B_u>$ & 
$<d_v^B> + 2<\widetilde{\lambda}^B_d >$ &
$<s_v^B> +  2<\widetilde{\lambda}^B_s>$ \\ \hline
&&& \vspace{-4mm}\\
$p$
 &$\widetilde{S}^3 + 2\widetilde{F} + \frac{2}{3}\widetilde{D}$
 &$\widetilde{S}^3 - \frac{4}{3}\widetilde{D}$
 &$\widetilde{S}^3 - 2\widetilde{F} + \frac{2}{3}\widetilde{D}$\\ 
$n$
 &$\widetilde{S}^3 - \frac{4}{3}\widetilde{D} $ 
 &$\widetilde{S}^3 + 2\widetilde{F} + \frac{2}{3}\widetilde{D} $ 
 &$\widetilde{S}^3 - 2\widetilde{F} +  \frac{2}{3}\widetilde{D} $ \\
$\Sigma^+ $
 &$\widetilde{S}^3 + \widetilde{F} + \frac{2}{3}\widetilde{D} $
 &$\widetilde{S}^3 - \widetilde{F} + \frac{2}{3}\widetilde{D} $
 &$\widetilde{S}^3 - \frac{4}{3}\widetilde{D} $ \\ 
$\Sigma^0 $ 
 &$\widetilde{S}^3 + \frac{2}{3}\widetilde{D} $
 &$\widetilde{S}^3 + \frac{2}{3}\widetilde{D} $
 &$\widetilde{S}^3 - \frac{4}{3}\widetilde{D} $ \\ 
$\Sigma^- $ 
 &$\widetilde{S}^3 - \widetilde{F} + \frac{2}{3}\widetilde{D}$
 &$ \widetilde{S}^3 + \widetilde{F} + \frac{2}{3}\widetilde{D}$
 &$\widetilde{S}^3 - \frac{4}{3}\widetilde{D}$ \\
$\Xi^{-}$ 
 &$\widetilde{S}^3 - 2\widetilde{F} + \frac{2}{3}\widetilde{D}$
 &$\widetilde{S}^3 - \frac{4}{3}\widetilde{D}$
 &$\widetilde{S}^3 + 2\widetilde{F} + \frac{2}{3}\widetilde{D}$ \\
$\Xi^{0}$
 &$\widetilde{S}^3 - \frac{4}{3}\widetilde{D}$
 &$\widetilde{S}^3 - 2\widetilde{F} + \frac{2}{3}\widetilde{D}$
 &$\widetilde{S}^3 + 2\widetilde{F} + \frac{2}{3}\widetilde{D}$ \\
$\Lambda^{0}$ 
 &$\widetilde{S}^3 - \frac{2}{3}\widetilde{D}$
 &$\widetilde{S}^3 - \frac{2}{3}\widetilde{D}$
 &$\widetilde{S}^3 + \frac{4}{3}\widetilde{D}$ \\
\end{tabular}
\end{table}
\begin{table}
\caption{The mean quantum number of the light sea quarks.
\label{T2}}
\begin{tabular}{llll}
 &$\langle I_3 \rangle$
 &$\langle Y \rangle$
 &$\langle Q \rangle$\\ 
$B$ &$\frac{1}{2}\{\langle \lambda_u^B - \lambda_d^B\rangle \}$ 
 &$\frac{1}{3}\{\langle \lambda_u^B + \lambda_d^B -2\lambda_s^B\rangle \}$
 &$\frac{1}{3}\{\langle 2\lambda_u^B - \lambda_d^B - \lambda_s^B\rangle \}$\\ \hline
&&& \vspace{-4mm}\\
$p$ 
 &$\frac{1}{2}(\widetilde{F} + \widetilde{D}) - \frac{1}{4}$
 &$\frac{1}{3}(3\widetilde{F} - \widetilde{D}) - \frac{1}{2}$
 &$\frac{1}{3}(3\widetilde{F} + \widetilde{D}) - \frac{1}{2}$ \\ 
 &$=-0.055$ &$ =0.56$ &$ =0.23$ \\ 
$n$ 
 &$-\frac{1}{2}(\widetilde{F} + \widetilde{D}) + \frac{1}{4}$
 &$\frac{1}{3}(3\widetilde{F} - \widetilde{D}) - \frac{1}{2}$
 &$-\frac{2}{3}\widetilde{D} $ \\
 &$=0.055$ & $=0.56$ & $=0.34$ \\  
$\Sigma^+$ 
 &$\frac{1}{2}\widetilde{F} - \frac{1}{2}$
 &$\frac{2}{3}\widetilde{D} $
 &$\frac{1}{6}(3\widetilde{F} + 2\widetilde{D}) - \frac{1}{2}$  \\
 &$=-0.054$ & $=-0.34$ & $=-0.22$ \\  
$\Sigma^0$ 
 &$0$
 &$\frac{2}{3}\widetilde{D} $
 &$\frac{1}{3}\widetilde{D}$ \\
 &$=0 $& $=-0.34$ &$ =-0.17$ \\ 
$\Sigma^-$ 
 &$-\frac{1}{2}\widetilde{F} + \frac{1}{2}$
 &$\frac{2}{3}\widetilde{D}$ 
 &$\frac{1}{6}(-3\widetilde{F} + 2\widetilde{D}) + \frac{1}{2}$ \\  
 &$=0.054$ & $=-0.34$ &$ =-0.11$ \\ 
$\Xi^-$ 
 &$\frac{1}{2}(-\widetilde{F} + \widetilde{D}) + \frac{1}{4}$
 &$-\frac{1}{3}(3\widetilde{F} + \widetilde{D}) + \frac{1}{2}$
 &$-\frac{1}{3}(3\widetilde{F} - \widetilde{D}) + \frac{1}{2} $\\ 
 &$=-0.45$ & $=-0.23$ & $=-0.56$ \\ 
$\Xi^0$ 
 &$\frac{1}{2}(\widetilde{F} - \widetilde{D}) - \frac{1}{4}$
 &$-\frac{1}{3}(3\widetilde{F} + \widetilde{D}) + \frac{1}{2}$
 &$-\frac{2}{3}\widetilde{D} $ \\ 
 &$=0.45$ & $=-0.23$ & $=0.34 $\\
$\Lambda^0 $
 &$0$
 &$-\frac{2}{3}\widetilde{D}$
 &$-\frac{1}{3}\widetilde{D}$ \\
 &$=0$ & $=0.34$ &$ =0.17$ \\
\end{tabular}
\end{table}
\begin{table}
\caption{The spin fraction of the quarks . 
\label{T3}}
\begin{tabular}{llll}
$B$ & $\triangle u^B$ & $\triangle d^B$ & $\triangle s^B$ \\ \hline
$p$
 &$\frac{1}{3}S + F + \frac{1}{3}D$
 &$\frac{1}{3}S - \frac{2}{3}D$
 &$\frac{1}{3}S - F + \frac{1}{3}D$\\ 
$n$
 &$\frac{1}{3}S - \frac{2}{3}D $ 
 &$\frac{1}{3}S + F + \frac{1}{3}D $ 
 &$\frac{1}{3}S - F +  \frac{1}{3}D $ \\
$\Sigma^+ $
 &$\frac{1}{3}S + \frac{1}{2}F + \frac{1}{3}D $
 &$\frac{1}{3}S - \frac{1}{2}F + \frac{1}{3}D $
 &$\frac{1}{3}S - \frac{2}{3}D $ \\ 
$\Sigma^0 $ 
 &$\frac{1}{3}S + \frac{1}{3}D $
 &$\frac{1}{3}S + \frac{1}{3}D $
 &$\frac{1}{3}S - \frac{2}{3}D $ \\ 
$\Sigma^- $ 
 &$\frac{1}{3}S - \frac{1}{2}F + \frac{1}{3}D$
 &$ \frac{1}{3}S + \frac{1}{2}F + \frac{1}{3}D$
 &$\frac{1}{3}S - \frac{2}{3}D$ \\
$\Xi^{-}$ 
 &$\frac{1}{3}S - F + \frac{1}{3}D$
 &$\frac{1}{3}S - \frac{2}{3}D$
 &$\frac{1}{3}S + F + \frac{1}{3}D$ \\
$\Xi^{0}$
 &$\frac{1}{3}S - \frac{2}{3}D$
 &$\frac{1}{3}S - F + \frac{1}{3}D$
 &$\frac{1}{3}S + F + \frac{1}{3}D$ \\
$\Lambda^{0}$ 
 &$\frac{1}{3}S - \frac{1}{3}D$
 &$\frac{1}{3}S - \frac{1}{3}D$
 &$\frac{1}{3}S + \frac{2}{3}D$ \\
\end{tabular}
\end{table}
\end{document}